\documentclass[pra,reprint,showpacs,floatfix,superscriptaddress]{revtex4-2}
\usepackage{bm}
\usepackage{placeins}
\usepackage[pdftex]{graphicx}
\usepackage{makecell,tabularx}
\usepackage{amsmath, amsthm, amssymb}

\newcommand{\footnoteremember}[2]{
\footnote{#2}
\newcounter{#1}
\setcounter{#1}{\value{footnote}}}
\newcommand{\footnoterecall}[1]{
\footnotemark[\value{#1}]}

\newcommand{\be}{\begin{eqnarray}}
\newcommand{\ee}{\end{eqnarray}}

\begin{document}

\title{Exact exchange-correlation potentials for calculating the fundamental gap with a fixed number of electrons}

\author{M. J. P. Hodgson}
\email[Personal email: ]{matthew.j.hodgson@durham.ac.uk}
\homepage[Personal webpage: ]{http://www-users.york.ac.uk/~mjph501/}
\affiliation{Department of Physics, Durham University, South Road, Durham, DH1 3LE, United Kingdom}
\affiliation{European Theoretical Spectroscopy Facility}

\author{J. Wetherell}
\email[Personal email: ]{jack.wetherell@polytechnique.edu}
\homepage[Personal webpage: ]{http://www-users.york.ac.uk/~jw1294/}
\affiliation{LSI, \'Ecole Polytechnique, CNRS, Institut Polytechnique de Paris, F-97728 Palaiseau}
\affiliation{European Theoretical Spectroscopy Facility}

\author{Emmanuel Fromager}
\affiliation{Laboratoire de Chimie Quantique, Institut de Chimie, CNRS / Universit\'{e} de Strasbourg, 4 rue Blaise Pascal, 67000 Strasbourg, France}

\begin{abstract}
Capturing the discontinuous shift by $\Delta$ in the exact exchange-correlation (xc) potential is the standard proposal for calculating the fundamental gap, $E_\mathrm{g}$, from the Kohn-Sham (KS) gap, $\varepsilon_\mathrm{g}$, within KS density functional theory (DFT), as $E_\mathrm{g} = \varepsilon_\mathrm{g} + \Delta$, yet this discontinuity is absent from existing approximations. The `$N$-centered' formulation of ensemble DFT artificially maintains a total electron number, $N$, in order to yield $E_\mathrm{g}$ not through a discontinuous shift in the xc potential but via the ensemble-weight derivative of the xc energy. Within the $N$-centered approach we calculate exact xc potentials for a one-dimensional finite system and show analytically that $\Delta$ can in fact be interpreted as a discontinuous shift in the exact $N$-centered ensemble xc potential, thereby extending to charged excitations an exact property of uncharged excitations. We show that applying the Levy-Zahariev `shift-in-potential' procedure in this context relocates the discontinuous shift to the unimportant periphery of the system, so that the exact xc potential in effect is free of discontinuities and thus the inability of a local functional to capture discontinuous behavior is inconsequential.
\end{abstract}

\maketitle

\section{Introduction}

Kohn-Sham \cite{PhysRev.140.A1133} (KS) density functional theory \cite{PhysRev.136.B864} (DFT) owes its success to its approximations, which achieve an unprecedented balance of computational efficiency and accuracy for electronic structure properties \cite{DG,LDA_Success2,burke2012perspective,becke2014perspective,giustino2014materials,jones2015density,kronik2016excited,verma2020status}. However, standard (semi-local) approximations are notoriously unreliable for predicting the fundamental gap \cite{mori2014derivative,baerends2017kohn}. As such, to predict reliable gaps, in conjunction with a KS calculation, computationally expensive methods outside of KS theory must be employed. These methods, which bypass the need for nonlocal dependence on the electron density within one's approximation to the exchange-correlation (xc) functional, use (more expensive) orbital-dependent functionals \cite{PhysRevB.99.045129}, e.g., hybrid density functionals \cite{perdew1996rationale,seidl1996generalized,salzner1997design,muscat2001prediction,heyd2003hybrid,paier2006screened,alkauskas2008defect,eisenberg2009new,jain2011reliability,PhysRevB.89.195134,crowley2016resolution,garza2016predicting,hait2016prediction,perdew2017understanding,kummel2017charge,PhysRevMaterials.2.040801}, which rely on a generalized KS scheme, or the $GW$ approximation within many-body perturbation theory \cite{PhysRev.139.A796,shishkin2007self,rinke2009defect,PhysRevB.81.085103,PhysRevLett.99.246403,gao2016speeding}. The extraction of accurate gaps solely from a regular KS calculation \cite{kraisler2014fundamental,verma2017hle16,verma2017hle17,guandalini2019fundamental,PhysRevB.101.115109} offers the prospect of improved semiconductor, thermoelectric material \cite{chen2016understanding,noor2016shift,jain2016computational,verma2020status} and photovoltaic \cite{chen2018density,trang2020theoretical,bainglass2020photovoltaic} modeling by drastically reducing the computational cost. 

In principle, KS DFT can yield the exact fundamental gap; in practice the exact xc energy of DFT and the corresponding xc potential of KS theory possess discontinuous features which have a strong nonlocal dependence on the electron density \cite{perdew1982density,almbladh1985density,PhysRevA.40.4190,:/content/aip/journal/jcp/131/22/10.1063/1.3271392,PhysRevA.83.062512,burke2012perspective,PhysRevB.93.155146,hodgson2017interatomic} and hence are difficult to capture in approximate functionals \cite{zhang1998challenge,:/content/aip/journal/jcp/125/20/10.1063/1.2403848,:/content/aip/journal/jcp/125/19/10.1063/1.2387954,sousa2007general,:/content/aip/journal/jcp/129/12/10.1063/1.2987202,cohen2008insights,mori2014derivative}. Perdew, Parr, Levy, and Balduz (PPLB) \cite{perdew1982density} proposed to calculate the gap -- the ionization potential (IP), $I$, minus the electron affinity (EA), $A$ -- within KS theory via the discontinuous constant shift, $\Delta$, in the multiplicative xc potential, $v_\mathrm{xc}(\bm{r})$, which occurs as the electron number, $\mathcal{N}$, infinitesimally surpasses an integer \cite{perdew1982density,kuisma2010kohn}, $N$, where
\begin{align} \label{Eq:DD}
    \Delta &= \lim_{\delta \rightarrow 0^+} \left ( \left. v_\mathrm{xc}(\bm{r}) \right |_{\mathcal{N}=N+\delta} - \left . v_\mathrm{xc}(\bm{r}) \right |_{\mathcal{N}=N-\delta} \right ) \nonumber \\
    &= I - A - \varepsilon_{N+1} + \varepsilon_N,
\end{align}
$\varepsilon_N$ is the KS highest occupied molecular orbital (HOMO) energy and $\varepsilon_{N+1}$ is the KS lowest unoccupied molecular orbital (LUMO) energy, both of the $N$-electron system. $\Delta$, usually termed the `derivative discontinuity', connects the fundamental gap to the KS gap, $\varepsilon_{N+1} - \varepsilon_N$. However, common density-functional approximations based on the local or semi-local density are unable to capture the discontinuous shift in the xc potential when the electron number infinitesimally surpasses an integer and hence do not yield $\Delta$ \cite{mardirossian2017thirty,PhysRevLett.77.3865}. 

Senjean and Fromager defined an `$N$-centered' KS system from which the exact fundamental gap can be calculated in principle \cite{senjean2018unified,senjean2020n}. Within their auxiliary system the electron number is fixed and integer with the purpose of calculating the system's fundamental gap from the derivative (at \textit{fixed} density) of the corresponding $N$-centered ensemble xc energy with respect to the ensemble weight rather than from a discontinuous shift in the xc potential. The theory was designed so that the exact xc potential does \textit{not} have to shift discontinuously in order to yield the gap and hence commonly used approximations to the xc energy, such as the local density approximation (LDA), could in principle be employed reliably in this context. 

In this paper we study the \textit{exact} $N$-centered approach: We calculate the exact ensemble xc energy as a function of the ensemble weight and the corresponding exact xc potentials for a one-dimensional finite model system for which the many-body Schr\"odinger equation can be solved exactly. We demonstrate both numerically and analytically that, when the exact ensemble xc potential is (arbitrarily) forced to asymptotically approach zero infinitely far from the center of the system, it experiences discontinuous shifts in the region of centralized density, just like in the PPLB approach. Importantly, we show how these shifts can be `relocated' from this central region of the system to the unimportant periphery via the exact Levy-Zahariev (LZ) shift \cite{levy2014ground} within the $N$-centered approach, even as the ensemble weight becomes infinitesimal. 

Our investigation into the nature of the $N$-centered method is organized as follows. For comparison, we first consider the ubiquitous PPLB approach; in Sec.~\ref{Sec:DD} we present a 1D model system which consists of two same-spin, interacting electrons in its ground state and calculate the exact KS xc potential upon the addition of a small fraction of an electron in order to observe the shift in the potential by $\Delta$. Next, in Sec.~\ref{Sec:N_cen}, we briefly review the established exact $N$-centered formalism. We then calculate, in Sec.~\ref{Sec:calculation_IP_and_EA}, charged excitation energies for our model system via the $N$-centered approach both exactly and employing an LDA: In Sec.~\ref{Sec:IP_exact} we calculate the IP and find the exact $N$-centered ensemble xc potential as a function of the ensemble weight. Section~\ref{Sec:IP_LDA} then employs a standard, \textit{weight-independent} LDA to the xc energy to find the IP for the same system, which highlights the importance of the derivative of the xc energy with respect to the ensemble weight. In Secs.~\ref{Sec:EA_exact} and \ref{Sec:EA_LDA} we repeat these calculations but now for the EA. In Sec.~\ref{Sec:NCDD} we present an analytic proof that $\Delta$ can indeed manifest within the exact $N$-centered ensemble xc potential as a discontinuous shift and thereby extend Levy's concept for neutral excitations to charged excitations \cite{PhysRevA.52.R4313,PhysRevA.90.042501}. We then summarize our work in Sec.~\ref{Sec:Conclusion}.  

\section{Perdew, Parr, Levy, and Balduz} \label{Sec:DD}

First let us study the PPLB approach in the absence of approximation. For all our numerical examples we employ our \texttt{iDEA} code \cite{PhysRevB.88.241102} which models electrons in 1D that interact via the appropriately softened Coulomb interaction $w_{\rm ee}(x,x')\equiv (\left | x - x' \right | + 1)^{-1}$ \cite{PhysRevA.72.063411}. The external potential for our system corresponds to a model 1D atom $v_\mathrm{ext}(x) = 3/(\left | x \right | + 1)$. Our system consists of two \textit{same-spin} electrons ($N=2$) and as such we can calculate the exact fully-correlated many-body wavefunction on a real-space grid in 1D and the many-body total energy. In order to find the exact many-body IP and EA of this system we also calculate the exact many-body wavefunction and energy of the anion (which consists of three same-spin electrons) and the wavefunction and energy of ion (i.e., the one-electron system). From these wavefunctions we can calculate the exact many-body electron density – both for the integer electron systems and as an ensemble. The exact ensemble many-body density is given by $n^\delta(x) = (1-\delta) n_N(x) + \delta n_{N+1}(x)$, where $n_i(x)$ is the many-body density of the $i$-electron system. From $n^\delta(x)$ we `reverse-engineer' the KS equations via the algorithm of Ref.~\onlinecite{PhysRevB.88.241102} to find the corresponding exact xc potential and KS energies for varying values of $\delta$.

\begin{figure}
    \centering
    \includegraphics[width=1.0\linewidth]{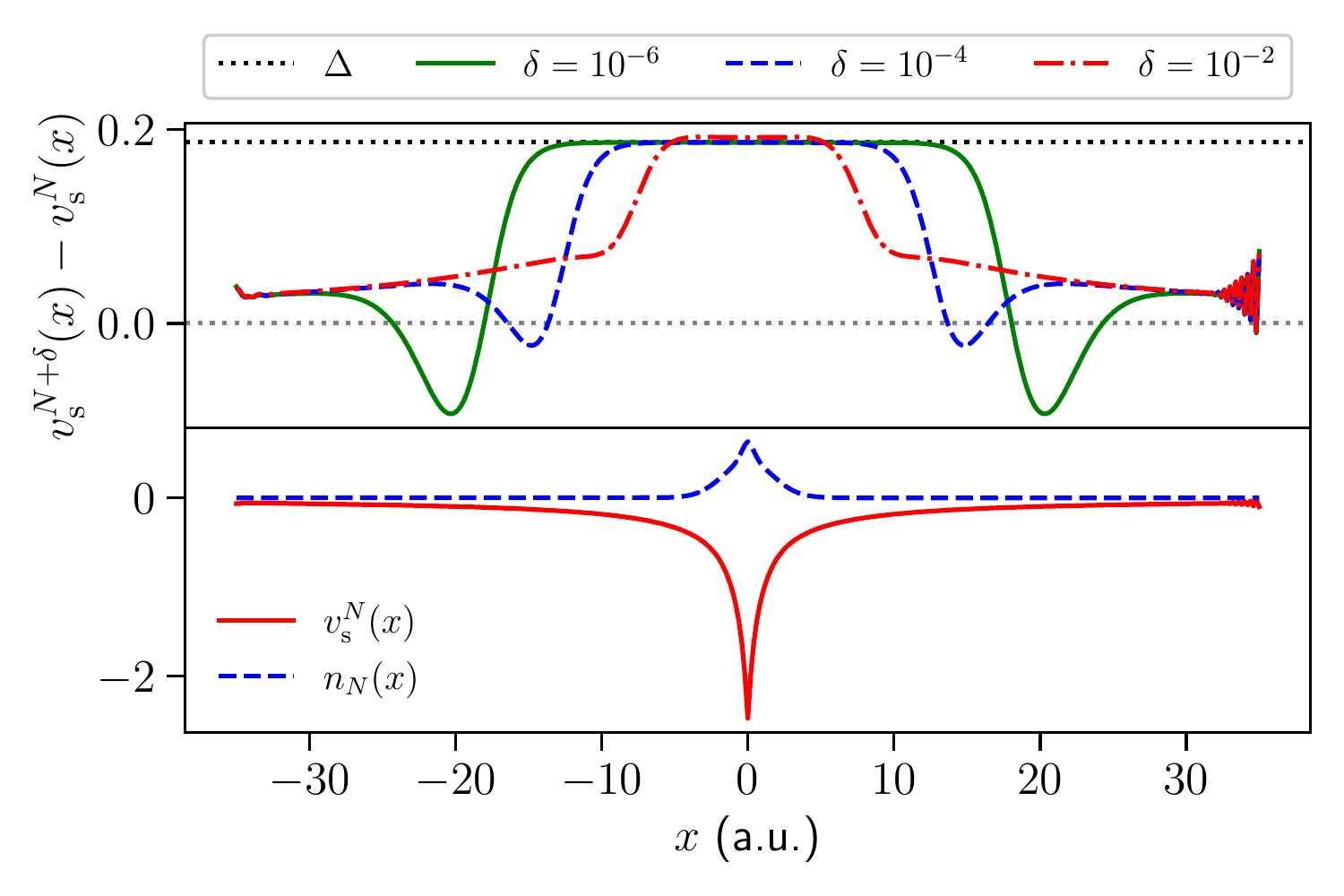}
    \caption{PPLB case: (Top) The change in the KS potential upon the addition of $\delta$ electrons to the system; as $\delta \rightarrow 0^+$ the change tends to a uniform shift of height $\Delta$ (numerical noise on the far right \footnote{We assume each xc potential decays $\propto -\frac{1}{|x|}$ and we align the overall constant shift of each potential such that $v_\mathrm{s}(|x| \rightarrow \infty)=0$ without modeling an infinitely large system.}). (Bottom) The 2-electron KS system to which $\delta$ electrons is added.}
    \label{Fig:PPLB}
\end{figure}

Employing the exact KS energies of the $N$-electron KS system, we calculate the exact $\Delta$ via Eq.~(\ref{Eq:DD}). Figure~\ref{Fig:PPLB} shows that as $\delta \rightarrow 0^+$ the change in the KS potential ($v^{N+\delta}_\mathrm{s}(x) - v^N_\mathrm{s}(x)$) \textit{tends} to a uniform constant of magnitude $\Delta$ \cite{hodgson2017interatomic,kraisler2020kohn}. In this case the $N$-electron KS potential is defined such that $v^{N+\delta}_\mathrm{s}(\left | x \right | \rightarrow \infty) = 0$. $v^{N+\delta}_\mathrm{s}(x)$ possess a discontinuous shift which elevates the potential in the \textit{central} region of the system. Note that as the plateau tends to a uniform constant shift, the KS potential still asymptotically approaches \textit{zero} infinitely far from the center of the system; as discussed in Refs.~\onlinecite{hodgson2017interatomic,kraisler2020kohn}.

This discontinuous behavior is difficult to capture in an approximate density functional and hence in practice the PPLB approach is not used. 

\section{$N$-centered ensemble approach} \label{Sec:N_cen}

We now turn to the $N$-centered approach. Following Ref.~\onlinecite{senjean2018unified}, we start from the general two-weight formulation of $N$-centered ensemble DFT where the ensemble density is
\begin{align} \label{Eq:Den}
    n^{\left\{\xi_-,\xi_+\right\}}(x)=&\left[1-\xi_-\frac{N-1}{N}-\xi_+\frac{N+1}{N}\right]n_N(x) \nonumber\\
    & +\xi_-n_{N-1}(x)+\xi_+n_{N+1}(x).
\end{align}
By analogy with Ref.~\onlinecite{senjean2020n}, we will refer to the
special cases $\left\{\xi_-,0\right\}$ and $\left\{0,\xi_+\right\}$ as \textit{left} and \textit{right} $N$-centered ensembles, respectively. In Sec.~\ref{Sec:calculation_IP_and_EA} we study in detail the left and then the right ensemble, for which we vary the ensemble weights as such: When $\xi_+ = 0$, $0 \leq \xi_- \leq \frac{N}{N-1}$, and when $\xi_- = 0$, $0 \leq \xi_+ \leq \frac{N}{N+1}$.\\

The general (two-weight) ensemble energy is
\begin{align}
    E^{\left\{\xi_-,\xi_+\right\}}=&\left[1-\xi_-\frac{N-1}{N}-\xi_+\frac{N+1}{N}\right]E_N \nonumber\\
    & +\xi_-E_{N-1}+\xi_+E_{N+1}.
\end{align}
The IP and EA can be extracted separately from the ensemble energy by differentiating with respect to $\xi_-$ and $\xi_+$; see Eqs.~(\ref{Eq:IP}) and (\ref{Eq:EA}). The ensemble energy, $E^{\left\{\xi_-,\xi_+\right\}}$, is a functional of the ensemble density $n^{\left\{\xi_-,\xi_+\right\}}$ which, unlike the PPLB approach, integrates to the fixed (`central') integral number of electrons, $N$. This many-body density is exactly reproduced by the non-interacting KS system:
\begin{align}\label{Eq:KSDen}
    &n^{\{\xi_-,\xi_+\}}(x) = \nonumber \\
    &\left ( 1-\xi_-\frac{N-1}{N}-\xi_+\frac{N+1}{N} \right ) \sum^N_{i=1}  \left | \phi^{\{\xi_-,\xi_+\}}_i(x) \right |^2 \nonumber \\ 
    &+ \xi_- \sum^{N-1}_{i=1} \left | \phi^{\{\xi_-,\xi_+\}}_i(x) \right |^2 + \xi_+ \sum^{N+1}_{i=1} \left | \phi^{\{\xi_-,\xi_+\}}_i(x) \right |^2,
\end{align}
where $\left \{ \phi^{\{\xi_-,\xi_+\}}_i(x) \right \}$ are the set of ensemble KS orbitals. As within the regular KS DFT approach, the auxiliary KS system employs the Hartree-xc (Hxc) potential, $v^{\left\{\xi_-,\xi_+\right\}}_\mathrm{Hxc}[n](x):=\delta E^{\left\{\xi_-,\xi_+\right\}}_{\rm Hxc}[n]/\delta n(x)$, to ensure that the non-interacting system yields the exact many-body ensemble density, where
\begin{equation}
    E^{\left\{\xi_-,\xi_+\right\}}_{\rm Hxc}[n]=E_{\rm H}[n]+E^{\left\{\xi_-,\xi_+\right\}}_{\rm xc}[n], \nonumber
\end{equation}
and $E_{\rm H}[n]$ is the conventional (weight-independent) Hartree functional. This formally exact deconstruction may not be optimal for approximate functionals as ghost-interaction errors \cite{gidopoulos2002spurious,pastorczak2014ensemble,PhysRevA.94.012511} may be introduced. Alternatively one may use exact (orbital-dependent) expressions for both Hartree and exchange ensemble energies \cite{loos2020weight,Gould2020_ugly} or opt for an alternative weight-dependent definition of the ensemble Hartree energy \cite{Gould2020_ensemble}. In this paper we mainly focus on the exact $N$-centered approach; for our LDA calculations we employ this deconstruction.

For brevity we adopt in the rest of this work the following shorthand notations when referring to ensemble densities, energies, and density functionals:
\be
\label{eq:notations_weights}
\begin{split}
\left\{\xi_-,0\right\} &\equiv \xi_-, \\\left\{0,\xi_+\right\} &\equiv \xi_+,
\\
\left\{\xi,\xi\right\} &\equiv \xi.
\end{split}
\ee
As a result, the left $N$-centered ensemble xc functional, for example, will simply be denoted as
\be
E^{\xi_-}_\mathrm{xc}[n]:=E^{\left\{\xi_-,0\right\}}_\mathrm{xc}[n].
\ee
Similarly, $E^{\xi_+}_\mathrm{xc}[n]:=E^{\left\{0,\xi_+\right\}}_\mathrm{xc}[n]$ and $E^{\xi}_\mathrm{xc}[n]:=E^{\left\{\xi,\xi\right\}}_\mathrm{xc}[n]$.

By design, the $N$-centered ensemble density integrates to $N$, which is fixed with the aim of conveniently expressing $\Delta=I-A-\varepsilon_{N+1}+\varepsilon_{N}$ as a weight derivative where the ensemble density is held constant \cite{senjean2018unified,senjean2020n}, as such
\begin{equation}\label{Eq:NCDD}
\begin{split}
    \Delta&=\left[\left.\dfrac{\partial E^{\xi_-}_\mathrm{xc}[n]}{\partial \xi_-}\right|_{\xi_-=0}+\left.\dfrac{\partial E^{\xi_+}_\mathrm{xc}[n]}{\partial \xi_+}\right|_{\xi_+=0}\right]_{n=n_N} \nonumber \\
    &\equiv\left.\dfrac{\partial E^{\xi}_\mathrm{xc}[n]}{\partial \xi}\right|_{\xi=0,n=n_N}.
\end{split}
\end{equation}
In Sec.~\ref{Sec:NCDD} we show analytically that $\Delta$ can also be interpreted, in the context of $N$-centered ensemble DFT, as a derivative discontinuity, exactly like in the PPLB approach.

\section{Extracting charged excitation energies}\label{Sec:calculation_IP_and_EA}

We now turn again to our model system and employ our exact many-body densities within the exact $N$-centered formalism in order to calculate charged excitation energies as a function of the ensemble weight. In addition we calculate the corresponding exact ensemble xc potentials and observe how they vary with the ensemble weight. 

Within this section we also employ a standard LDA to the ensemble xc energy which completely neglects the ensemble-weight dependence. 

\subsection{Ionization potential} \label{Sec:IP}

\subsubsection{Exact theory} \label{Sec:IP_exact}

We start from the general expression of the IP for an $N$-electron system within the $N$-centered approach -- Eq.~(33) in Ref.~\onlinecite{senjean2018unified}. By considering the left $N$-centered ensemble we obtain
\begin{equation} \label{Eq:IP}
    I = -\varepsilon^{\xi_-}_N +\left[ \left(\dfrac{\xi_-}{N}+1\right)\dfrac{\partial E^{\xi_-}_\mathrm{xc}[n]}{\partial \xi_-} -C^{\xi_-}[n] \right]_{n=n^{\xi_-}},
\end{equation}
where $C^{\xi_-}[n]:=C^{\left\{\xi_-,0\right\}}[n]$ is the analog of the LZ shift in potential for $N$-centered ensembles \cite{levy2014ground} (note that the notations of Eq.~(\ref{eq:notations_weights}) have been employed). The general two-weight expression for the ensemble density-functional LZ shift is
\begin{equation} \label{Eq:LZ}
    C^{\left\{\xi_-,\xi_+\right\}}[n] = \frac{E^{\left\{\xi_-,\xi_+\right\}}_\mathrm{Hxc}[n] - \int \mathrm{d} x \ n(x) v^{\left\{\xi_-,\xi_+\right\}}_\mathrm{Hxc}[n](x)}{\int \mathrm{d}x \ n(x)}.
\end{equation}
We stress that, unlike in the conventional PPLB approach, the HOMO and LUMO energies ($\varepsilon^{\xi_-}_N$ and $\varepsilon^{\xi_+}_{N+1}$, respectively) are defined \textit{up to an additive constant}. This is connected to the fact that the left and right $N$-centered ensemble densities ($n^{\xi_-}$ and $n^{\xi_+}$, respectively) integrate to the (fixed and integer) central number of electrons, $N$. On the other hand, the LZ-shifted energies, $\overline{\varepsilon}^{\xi_-}_N=\varepsilon^{\xi_-}_N+\left.C^{\xi_-}[n]\right|_{n=n^{\xi_-}}$ and $\overline{\varepsilon}^{\xi_+}_{N+1}= \varepsilon^{\xi_+}_{N+1}+\left.C^{\xi_+}[n]\right|_{n=n^{\xi_+}}$, are \textit{truly unique} and independent of any overall shift in the potential.

\begin{figure}
    \centering
    \includegraphics[width=1.0\linewidth]{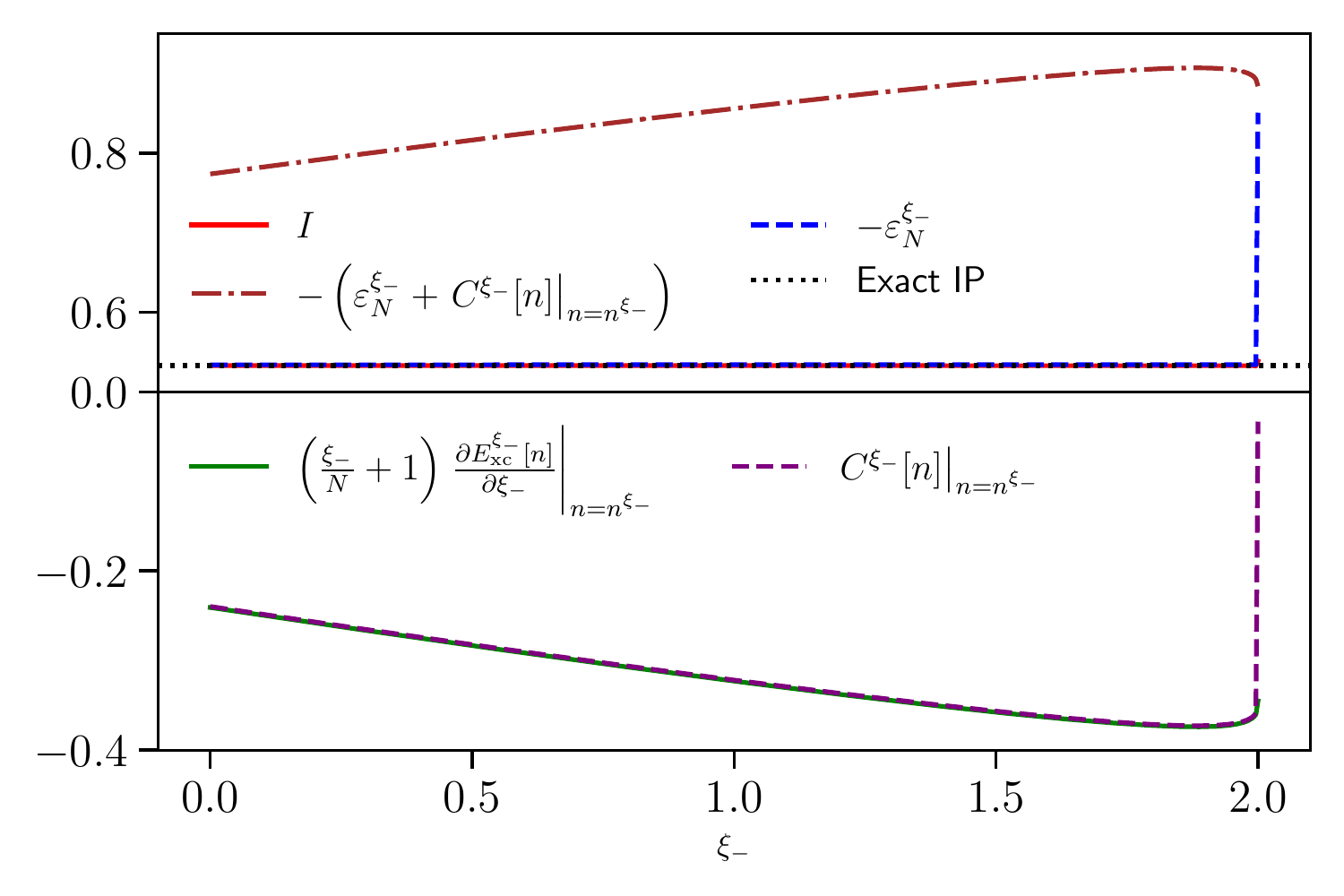}
    \caption{The exact KS HOMO energy ($\varepsilon^{\xi_-}_N$) is minus the exact IP for $0 \leq \xi_- < \frac{N}{N-1}$. Each term of Eq.~(\ref{Eq:IP}) is shown. The calculation of $I$ via Eq.~(\ref{Eq:IP}) equals the exact IP for all $\xi_-$ as expected. We choose $v^{\xi_-}_\mathrm{s}(x)$ to asymptotically approach 0.}
    \label{Fig:IP_exact}
\end{figure}

We study the same 1D atom of Sec.~\ref{Sec:DD} but now the electron number is fixed at $N=2$ within the $N$-centered approach. First we compare the exact IP to that of Eq.~(\ref{Eq:IP}). We calculate the exact many-body ensemble density via Eq.~(\ref{Eq:Den}) considering only the left contribution and employing the exact many-body density for the $N$- and ($N-1$)-electron systems. We then employ the ensemble density, $n^{\xi_-}(x)$, and reverse-engineer the KS equations to find the exact ensemble xc potential and KS energies. Our results are shown in Fig.~\ref{Fig:IP_exact}. We choose the ensemble KS potential that tends to zero as $\left | x \right | \rightarrow \infty$.

With this choice of KS potential we find that for $0 \leq \xi_- < \frac{N}{N-1}$, $I = -\varepsilon^{\xi_-}_N$ and hence 
\begin{equation}\label{Eq:IPXCC}
    \left . \left ( \frac{\xi_-}{N} + 1 \right ) \frac{\partial E^{\xi_-}_\mathrm{xc}[n]}{\partial \xi_-} \right |_{n = n^{\xi_-}} = \left . C^{\xi_-}[n] \right |_{n = n^{\xi_-}}.
\end{equation}
This can be proven analytically, as follows: For $0 \leq \xi_- < \frac{N}{N-1}$ the $n_N(x)$ term dominates the asymptotic decay of the density (Eq.~(\ref{Eq:Den})) far from the center of the system:
\begin{equation} \label{Eq:ADD}
    \lim_{\left | x \right | \rightarrow \infty} n^{\xi_-}(x) = \left ( 1 - \frac{N-1}{N} \xi_- \right ) n_N(x) \propto e^{-2 \sqrt{2 I}\left | x \right |}.
\end{equation}
In addition, in the asymptotic region, the KS orbital with the highest energy ($\varepsilon^{\xi_-}_N$) will dominate the KS density (Eq.~(\ref{Eq:KSDen})): 
\begin{align}
    \lim_{\left | x \right | \rightarrow \infty} n^{\xi_-}(x) &= \left ( 1 - \frac{N-1}{N} \xi_- \right ) \left | \phi^{\xi_-}_N(x) \right |^2 \nonumber \\
    &\propto e^{-2 \sqrt{-2 \varepsilon^{\xi_-}_N}\left | x \right |}, \nonumber
\end{align}
as, by design, $v^{\xi_-}_\mathrm{s}(\left | x \right | \rightarrow \infty) = 0$. Hence, as this exponential decay equals that of Eq.~(\ref{Eq:ADD}) by definition, $I = -\varepsilon^{\xi_-}_N$ and from Eq.~(\ref{Eq:IP}), Eq.~(\ref{Eq:IPXCC}) must follow.

We now examine $v^{\xi_-}_\mathrm{xc}[n^{\xi_-}](x) \equiv v^{\xi_-}_\mathrm{xc}(x)$ and  $v^{\xi_-}_\mathrm{xc}(x)+\left.C^{\xi_-}[n]\right|_{n=n^{\xi_-}}$ as $\xi_- \rightarrow 2^-$ $\left ( \frac{N}{N-1}=2 \right )$ which is the fully ionized limit, i.e., $n^{\xi_-}\overset{\xi_-=\frac{N}{N-1}}{\longrightarrow}\frac{N}{N-1}n_{N-1}$ to see if there is any discontinuity in the exact xc potential. 

\begin{figure}
    \centering
    \includegraphics[width=1.0\linewidth]{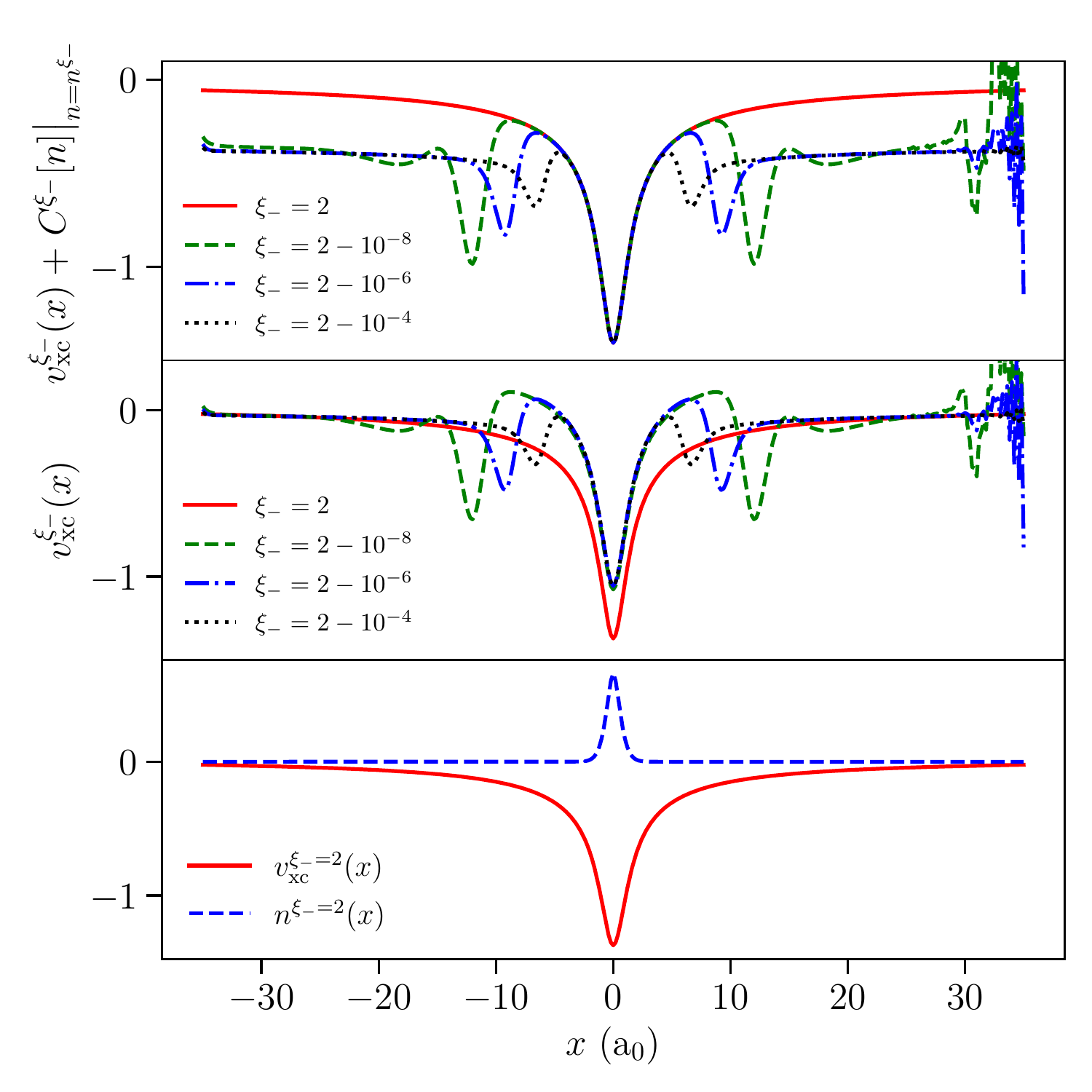}
    \caption{(Top) The exact xc potential plus the LZ shift. There is no discontinuous shift in the potential as $\xi_-\rightarrow 2^-$ in the region of centralized density (numerical noise on the far right). (Middle) The exact xc potential. There is a discontinuous shift in the potential as $\xi_- \rightarrow 2^-$. (Bottom) The density and xc potential for $\xi_-=2$.}
    \label{Fig:IP_exact_atom}
\end{figure}

Figure~\ref{Fig:IP_exact_atom} (Top) shows that with the LZ shift there is no plateau which yields a discontinuous shift in the xc potential in the region of the atom and hence approximate functionals which do not capture such discontinuities could be employed as reliable approximations to $v^{\xi_-}_\mathrm{xc}(x)+\left.C^{\xi_-}[n]\right|_{n=n^{\xi_-}}$, unlike within the PPLB approach. On the other hand, Fig.~\ref{Fig:IP_exact_atom} (Middle) shows $v^{\xi_-}_\mathrm{xc}$ which contains a plateau similar to that of Fig.~\ref{Fig:PPLB}. The LZ shift `relocates' the discontinuity from the centralized region to the edges of the system -- a region with little influence on the system's properties \footnoteremember{SM}{A video of the exact $N$-centered ensemble potentials and density as a function of the ensemble weight can be found in our Supplemental Material at XXXXX.}. As $\xi_- \rightarrow 2^-$, this plateau would become infinitely expansive, exactly like for the PPLB case; see Fig.~\ref{Fig:PPLB}. 

Figure~\ref{Fig:IP_exact} clearly shows the importance of the ensemble-weight dependence in the xc energy when we employ the `discontinuity-free' LZ-shifted xc potential. Without the contribution from the ensemble xc energy, the LZ shift significantly worsens the IP predicted from Eq.~(\ref{Eq:IP}). Therefore, if one wishes to remove discontinuous shifts from the exact xc potential, one requires a reliable approximation to the ensemble-weight dependence in the xc energy within the $N$-centered approach. 

\subsubsection{The local density approximation} \label{Sec:IP_LDA}

Next we calculate the IP within the $N$-centered approach employing the LDA to the xc energy, which neglects the weight dependence entirely. Our LDA to the xc energy was parameterized from the 1D homogeneous electron gas \cite{PhysRevB.97.235143}. The LDA does not capture PPLB's discontinuous shift, $\Delta$, in the xc potential. Therefore, although it is unsuitable for predicting the gap within the PPLB approach, it can be employed within the $N$-centered approach.

\begin{figure}
    \centering
    \includegraphics[width=1.0\linewidth]{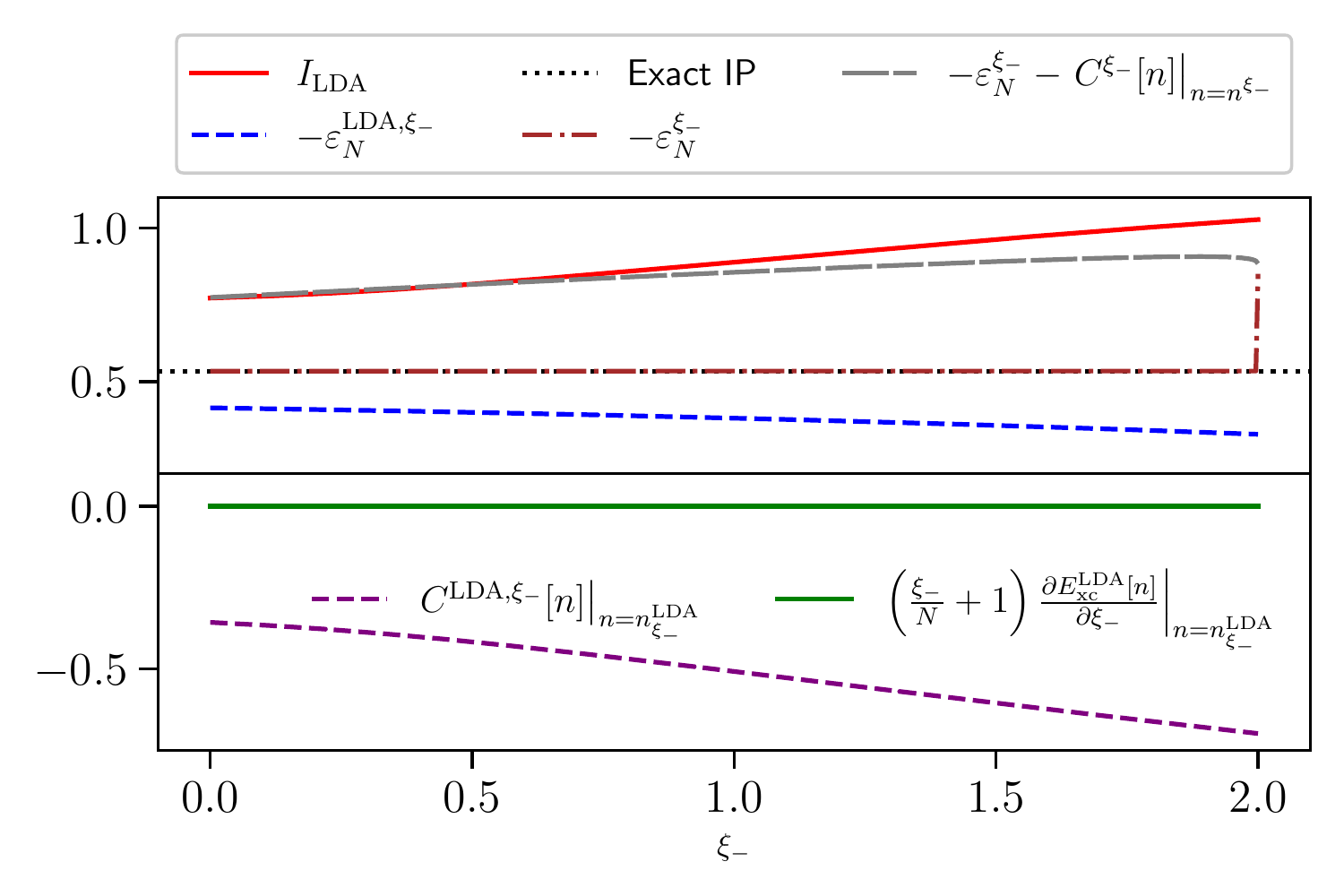}
    \caption{Same as Fig.~\ref{Fig:IP_exact} but an LDA to the xc energy has been employed. $I^\mathrm{LDA}$ is approximately equal to the exact $-\varepsilon_N^{\xi_-}-\left . C^{\xi_-}[n] \right \vert_{n=n^{\xi_-}}$ for approximately $0 \leq \xi^-<1.0$ showing the importance of the xc energy's ensemble-weight dependence in obtaining an accurate IP. $\varepsilon_N^{\mathrm{LDA},\xi_-}$ is inaccurate and hence a cancellation of errors with the LDA LZ shift must occur.}
    \label{Fig:IP_LDA}
\end{figure}

Figure~\ref{Fig:IP_LDA} shows the prediction for the IP evaluated via Eq.~(\ref{Eq:IP}) within the LDA. $I^\mathrm{LDA}$ is comparable to the exact LZ-shifted HOMO KS energy. Therefore, if an accurate approximation to the weight derivative of the xc energy were introduced, the LDA could yield a reliable prediction for $I$. This highlights the importance of developing reliable approximations to the xc energy which account for the ensemble-weight dependence \cite{loos2020weight,marut2020weight}.

\begin{figure}
    \centering
    \includegraphics[width=1.0\linewidth]{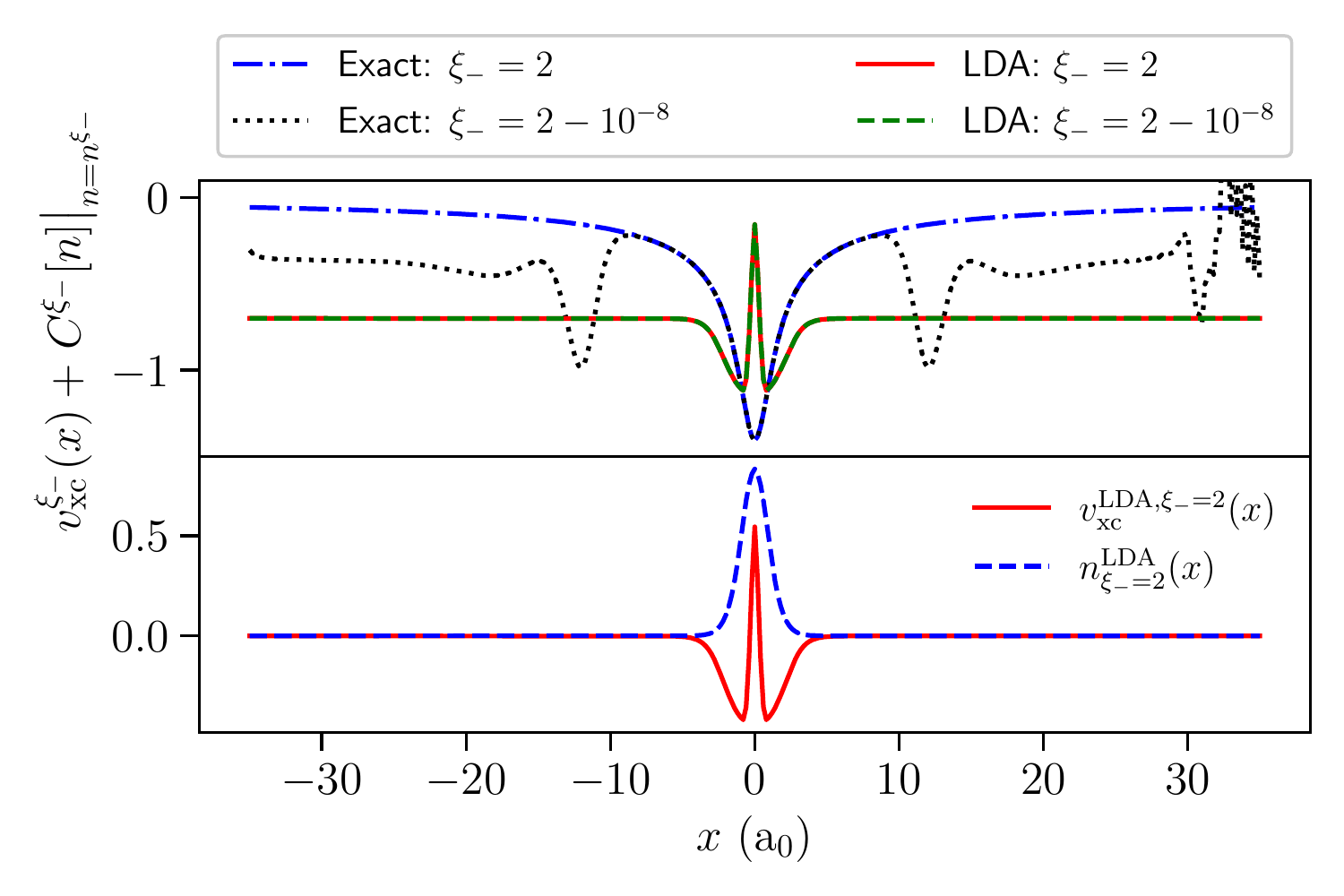}
    \caption{The LDA xc potential for $\xi_- \rightarrow 2^-$ and $\xi_- = 2$. The LDA xc potential does not visibly change as $\xi_- \rightarrow 2$. The exact potential is shown for reference.}
    \label{Fig:IP_LDA_atom}
\end{figure}

We see in Fig.~\ref{Fig:IP_LDA} that the LDA reproduces the LZ-shifted HOMO energy well, especially when $\xi_-\leq 1$. However, Fig.~\ref{Fig:IP_LDA_atom} shows that when $\xi_- = 2$, the LDA LZ-shifted xc potential is inaccurate, including the LDA LZ shift. This is likely because for $\xi_- > 1$ this LDA is employed for an ensemble well beyond the ground state systems from which it was parameterized. Despite this the error in the LDA LZ-shifted HOMO energy is acceptable owing to a cancellation of errors between the approximate xc energy (which contributes to the LZ shift) and the LZ-shifted potential which is overestimated in the region of the atom and underestimated away from it.

\subsection{Electron affinity}

\subsubsection{Exact theory} \label{Sec:EA_exact}

We now turn to the EA. We employ the right ensemble system for which $\xi_+$ is the ensemble weight; see Eq.~(\ref{Eq:Den}). The exact EA for the $N$-electron many-body system is given by the following expression~\cite{senjean2018unified}:
\begin{equation} \label{Eq:EA}
    A = -\varepsilon^{\xi_+}_{N+1} + \left [ \left ( \frac{\xi_+}{N} - 1 \right ) \frac{\partial E^{\xi_+}_\mathrm{xc}[n]}{\partial \xi_+} - C^{\xi_+}[n] \right ]_{n = n^{\xi_+}}.
\end{equation}

\begin{figure}
    \centering
    \includegraphics[width=1.0\linewidth]{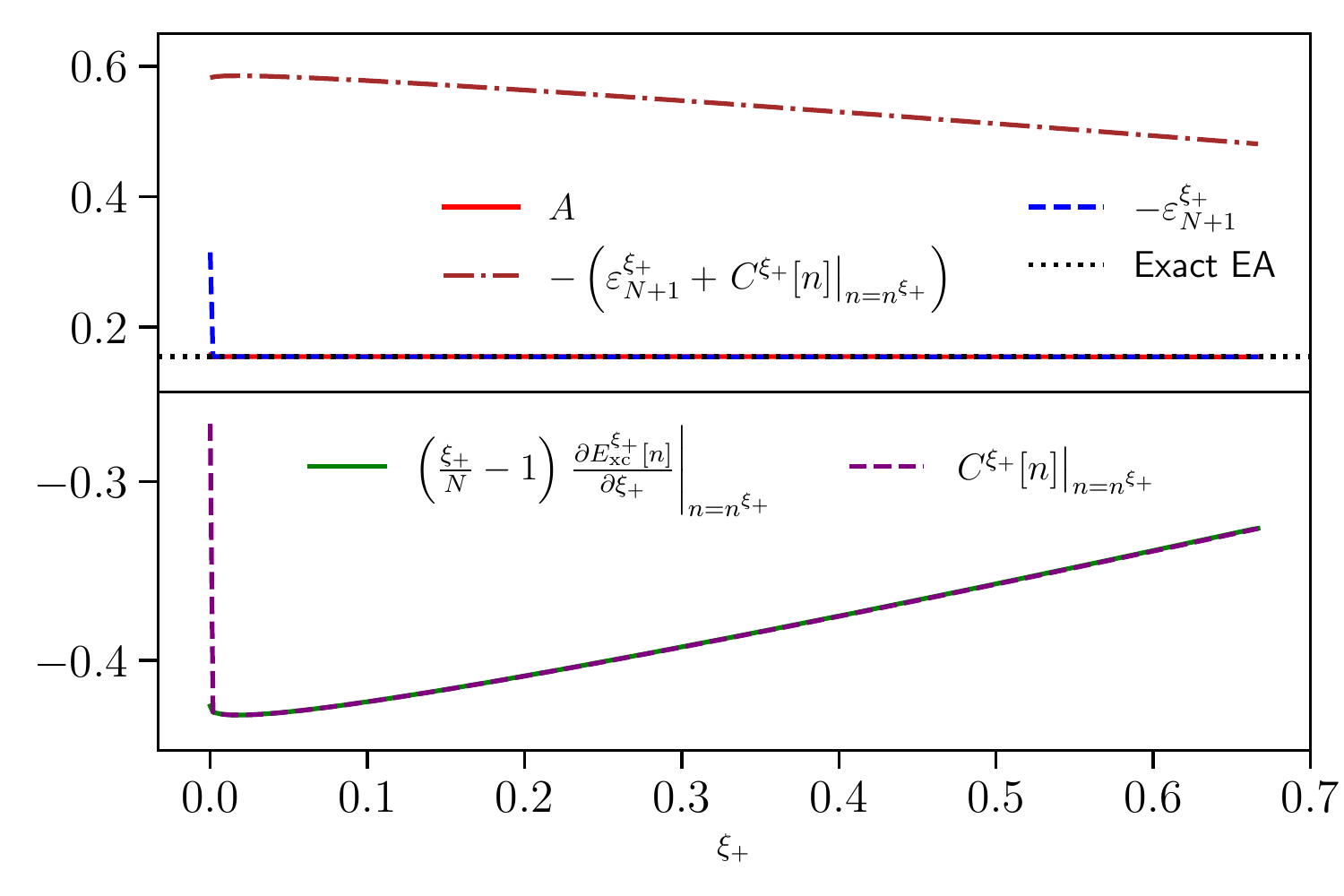}
    \caption{The exact KS LUMO energy ($\varepsilon^{\xi_+}_{N+1}$) is minus the exact EA for $0 < \xi_+ \leq \frac{N}{N+1}$. The separated terms of Eq.~(\ref{Eq:EA}) are also shown. The calculation of $A$ via Eq.~(\ref{Eq:EA}) equals the exact EA for all $\xi_+$, as expected. We choose $v^{\xi_+}_\mathrm{s}(x)$ to asymptotically approach 0.}
    \label{Fig:EA_exact}
\end{figure}

We model the same 1D atom as above. As before, we calculate the exact many-body ensemble density via Eq.~(\ref{Eq:Den}), with $\xi_- = 0$ and $\xi_+$ allowed to vary, employing the exact many-body density for the $N$- and ($N+1$)-electron systems. We then employ $n^{\xi_+}(x)$ and reverse-engineer the KS equations to find the exact ensemble xc potential and KS energies. Our results are shown in Fig.~\ref{Fig:EA_exact} \footnoterecall{SM}: We compare our calculated EA to the exact obtained via total energy differences of the $N$- and ($N+1$)-electron systems. As before, we (arbitrarily) choose a KS potential, $v^{\xi_+}_\mathrm{s}(x)$, that asymptotically approaches 0 towards infinity. Note that the expression on the right-hand side of Eq.~(\ref{Eq:EA}) is invariant under a constant shift in the KS potential, by design.

Like for the IP, $\varepsilon^{\xi_+}_{N+1}$ is found to be independent of $\xi_+$ for $0<\xi_+\leq \frac{N}{N+1}$. This is because the exact many-body ensemble density decays as 
\begin{equation}
    \lim_{\left | x \right | \rightarrow \infty} n^{\xi_+}(x) = \xi_+ n_{N+1}(x) \propto e^{-2 \sqrt{2 A}\left | x \right |} \nonumber
\end{equation}
far from the center of the system for $0 < \xi_+ \leq \frac{N}{N+1}$. In this asymptotic region, the KS orbital with the highest energy ($\varepsilon^{\xi_+}_{N+1}$) dominates the KS density provided $\xi_+ > 0$, and hence
\begin{equation}
    \lim_{\left | x \right | \rightarrow \infty} n^{\xi_+}(x) = \xi_+ \left | \phi^{\xi_+}_{N+1}(x) \right |^2 \propto e^{-2 \sqrt{- 2 \varepsilon^{\xi_+}_{N+1} } \left | x \right |}. \nonumber
\end{equation}
Therefore, as the many-body ensemble density and the KS density are the same, $A = -\varepsilon^{\xi_+}_{N+1}$ and 
\begin{equation}
    \left . \left ( \frac{\xi_+}{N} - 1 \right ) \frac{\partial E^{\xi_+}_\mathrm{xc}[n]}{\partial \xi_+} \right |_{n = n^{\xi_+}} = \left . C^{\xi_+}[n] \right |_{n = n^{\xi_+}}, \nonumber
\end{equation}
for $0 < \xi_+ \leq  \frac{N}{N+1}$.

\begin{figure}
    \centering
    \includegraphics[width=1.0\linewidth]{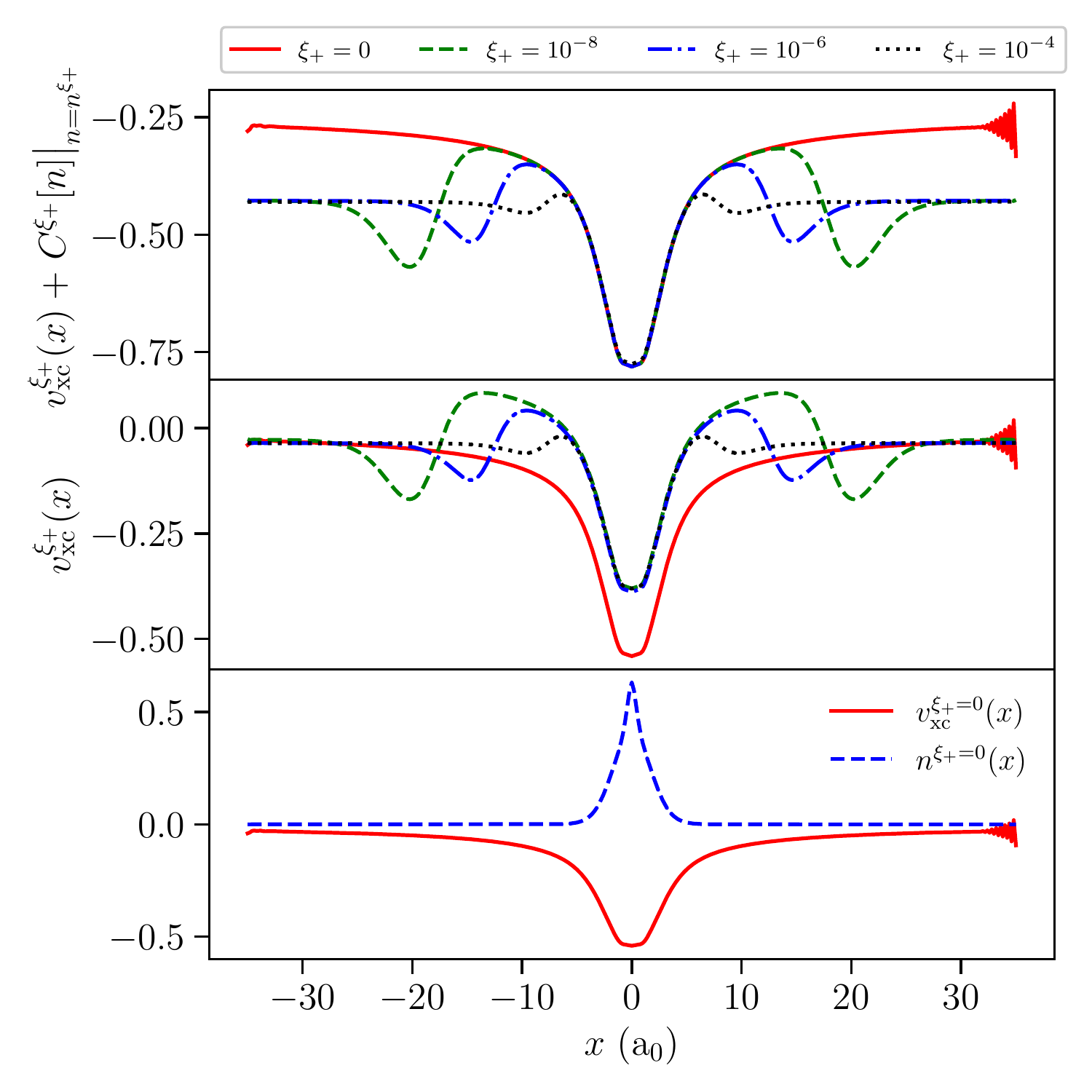}
    \caption{(Top) The exact xc potential plus the LZ shift. There is no discontinuous shift in the potential as $\xi_+ \rightarrow 0^+$ (numerical noise on the far right). (Middle) The exact xc potential. There is a discontinuous shift in the potential as $\xi_+ \rightarrow 0^+$. (Bottom) The density and xc potential for $\xi_+=0$. We choose $v^{\xi_+}_\mathrm{s}(x)$ to asymptotically approach 0.}
    \label{Fig:EA_exact_atom}
\end{figure}

We now examine $v^{\xi_+}_\mathrm{xc}$ and $v^{\xi_+}_\mathrm{xc}+\left . C^{\xi_+}[n] \right \vert_{n=n^{\xi^+}}$ as $\xi_+ \rightarrow 0^+$ to observe if in this case there is a discontinuous shift in the potential. As for the IP in Sec.~\ref{Sec:IP}, we observe no shift in the potential in the region of the centralized electron density when the LZ shift is included, as expected \cite{senjean2020n}; see Fig.~\ref{Fig:EA_exact_atom} (Top). Instead the shift occurs in the outer region of the atom and as $\xi_+ \rightarrow 0^+$ the steps we see in Fig.~\ref{Fig:EA_exact_atom} (Top) move further and further from the atom towards infinity. Without the LZ shift we observe a shift in the potential in the vicinity of the atom (Fig.~\ref{Fig:EA_exact_atom} (Middle)), as for the IP case above. In this case it can be shown that the height of the steps equals $\Delta$ as $\xi_+ \rightarrow 0^+$; see Sec.~\ref{Sec:NCDD}.

\subsubsection{The local density approximation} \label{Sec:EA_LDA}

\begin{figure}
    \centering
    \includegraphics[width=1.0\linewidth]{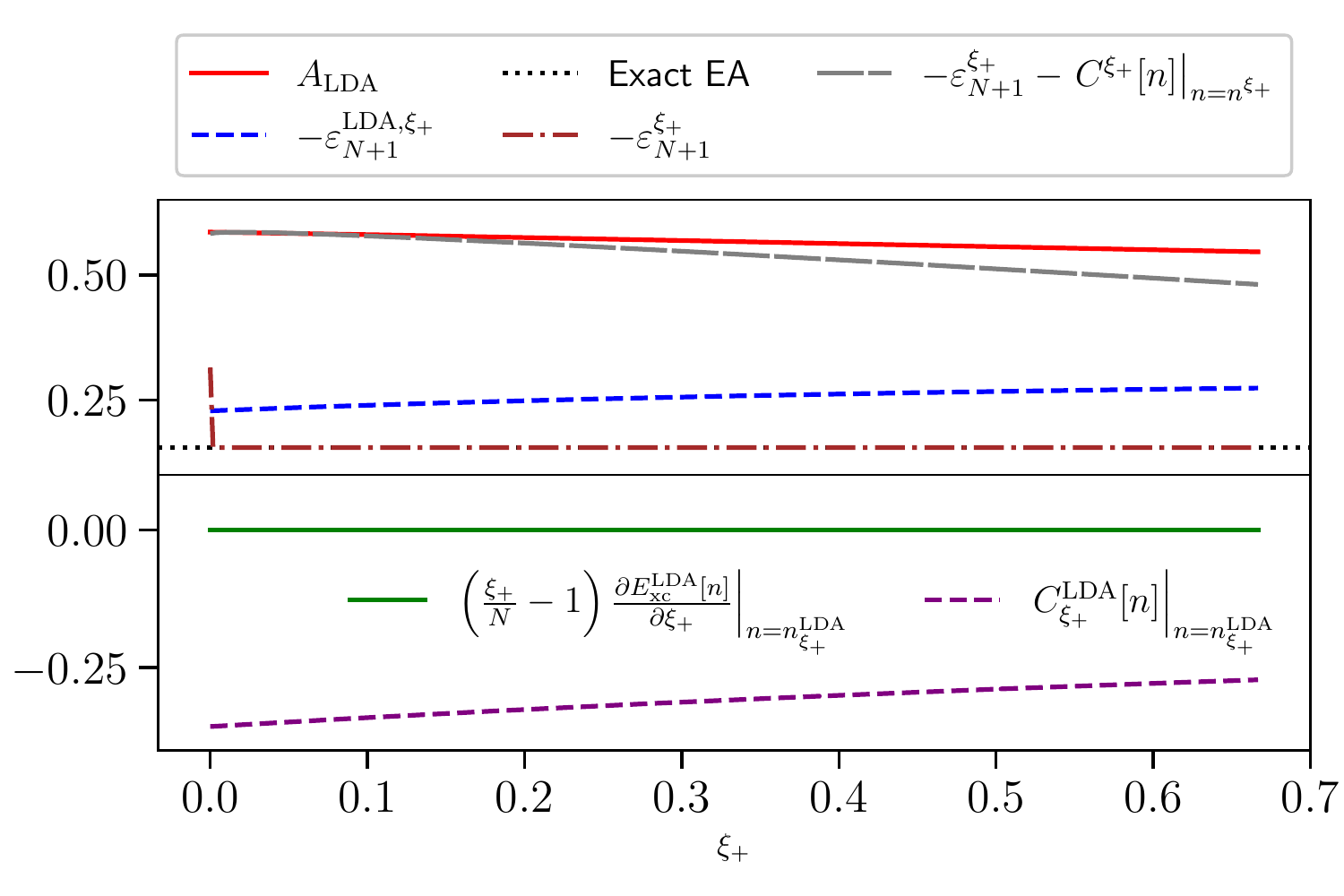}
    \caption{Same as Fig.~\ref{Fig:EA_exact} but an LDA to the xc energy has been employed. The ensemble-weight dependence is clearly crucial for obtaining an accurate EA. For reference the exact $\varepsilon_{N+1}^{\xi_+}$ is shown.}
    \label{Fig:EA_LDA}
\end{figure}

Figure~\ref{Fig:EA_LDA} shows the LDA's prediction for the EA. Once again it is clear that the ensemble-weight dependence is important for the EA prediction within the $N$-centered system; see Sec.~\ref{Sec:IP_LDA}.

\begin{figure}
    \centering
    \includegraphics[width=1.0\linewidth]{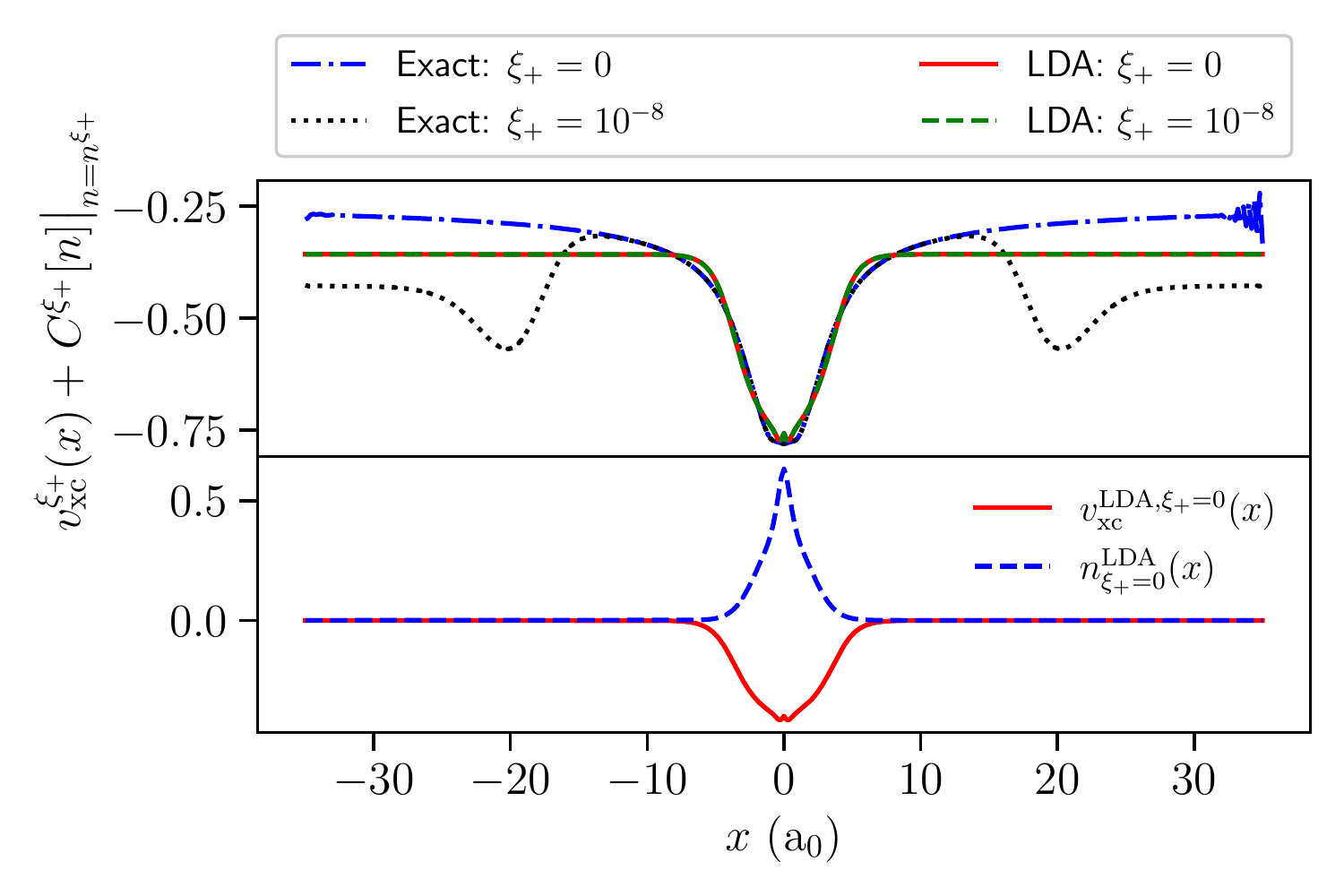}
    \caption{The LDA xc potential as $\xi_+ \rightarrow 0^+$ and $\xi_+=0$. The LDA xc potential does not visibly change as $\xi_+ \rightarrow 0$. The exact xc potential is shown for reference.}
    \label{Fig:EA_LDA_atom}
\end{figure}

As for the IP, the LDA yields an accurate LZ-shifted KS energy for $\xi_+<0.5$; see Fig.~\ref{Fig:EA_LDA}. Figure~\ref{Fig:EA_LDA_atom} shows that the LDA xc potential decays too quickly (a standard issue of the LDA) which yields an error in $\varepsilon^{\mathrm{LDA},\xi_+}_{N+1}$, however, this error is cancelled by the LDA LZ shift, as in Sec.~\ref{Sec:IP_LDA}. Therefore, in this case, with the addition of a reliable approximation to the ensemble-weight derivative of the xc energy, one could obtain an accurate EA from the LDA within the $N$-centered approach.

For both the left and the right ensemble systems as $\xi \rightarrow 0$ the LDA LZ-shifted KS energies are reliable approximations to the exact LZ-shifted KS energies for the $N$-electron system owing to a cancellation of errors between the LDA KS energies and the LDA LZ shift; see Figs.~\ref{Fig:IP_LDA}, \ref{Fig:IP_LDA_atom} and \ref{Fig:EA_LDA}. Further investigation is required to determine whether this cancellation of errors occurs for other types of system. These results imply that with an accurate local approximation to the ensemble-weight derivative of the xc energy, in the spirit of previous works on neutral excitations \cite{loos2020weight,marut2020weight}, the $N$-centered approach within the LDA could yield accurate fundamental gaps for a low computational cost.

\section{`Left-to-right' discontinuous shift in the $N$-centered exchange-correlation potential} \label{Sec:NCDD}

In the original formulation of $N$-centered ensemble DFT~\cite{senjean2018unified}, it was emphasized that modeling derivative discontinuities in order to recover the fundamental gap from the KS gap is unnecessary. Instead, one should focus on the weight dependence of the ensemble xc energy. Still, in the light of our numerical results and by analogy with Gross-Oliveira-Kohn DFT \cite{gross1988density,oliveira1988density,gross1988rayleigh,PhysRevA.52.R4313} for neutral excitations, one may wonder if a clear and formal connection can be established between weight derivatives and derivative discontinuities in the context of $N$-centered ensemble DFT. Below we establish this connection.

The asymptotic behavior of the left and right $N$-centered ensemble densities revealed that $I = -\left(\varepsilon^{\xi_-}_N-v^{\xi_-}_\mathrm{xc}(\left | x \right | \rightarrow \infty)\right)$, for $0\leq\xi_-<\frac{N}{N-1}$, and $A = -\left(\varepsilon^{\xi_+}_{N+1}-v^{\xi_+}_\mathrm{xc}(\left | x \right | \rightarrow \infty)\right)$, for $0<\xi_+\leq \frac{N}{N+1}$. In the PPLB approach, the xc potential tends to zero infinitely far from the center of the system, as a consequence of Janak's theorem. In the $N$-centered picture, the potential is unique up to a constant. If we make the (arbitrary) choice that the $N$-centered ensemble xc potential always tends to zero at infinite distance,
\begin{equation}\label{eq:pot_infinity_constr}
    v^{\xi_-}_\mathrm{xc}(\left | x\right | \rightarrow \infty)=v^{\xi_+}_\mathrm{xc}(\left | x\right | \rightarrow \infty)=0,
\end{equation}
then the LZ shift and weight derivative contributions to the IP/EA
compensate each other exactly (shown above), which can be written more
explicitly as follows:  
\begin{align}\label{eq:compensation_LZ_DD}
    E^{\xi_\pm}_\mathrm{Hxc}\left[n^{\xi_\pm}\right] -\int \mathrm{d}x\, &v^{\xi_\pm}_\mathrm{Hxc}(x) n^{\xi_\pm}(x) = \nonumber \\
    &\left(\xi_\pm\mp N\right)\left.\dfrac{\partial E^{\xi_\pm}_\mathrm{xc}[n]}{\partial \xi_\pm}\right|_{n=n^{\xi_\pm}}.
\end{align}
Note that, with the constraint in Eq.~(\ref{eq:pot_infinity_constr}), $v^{\xi_-=0}_\mathrm{xc}(x)=v^{\xi_+=0}_\mathrm{xc}(x)$. As a result, by considering both $\xi_-=0$ and $\xi_+\rightarrow 0^+$ cases in Eq.~(\ref{eq:compensation_LZ_DD}), we obtain by subtraction   
\begin{equation}\label{eq:connection_weightderiv_DD}
\begin{split}
    &\int \mathrm{d}x\left(v^{\xi_+\rightarrow 0^+}_\mathrm{xc}(x)-v^{\xi_-=0}_\mathrm{xc}(x)\right)n_{N}(x) \\
    &=\int \mathrm{d}x\left(v^{\xi_+\rightarrow 0^+}_\mathrm{xc}(x)-v^{\xi_+=0}_\mathrm{xc}(x)\right)n_{N}(x) \\
    &=N\left(\left.\dfrac{\partial E^{\xi_+}_\mathrm{xc}[n_N]}{\partial \xi_+}\right|_{\xi_+=0} + \left.\dfrac{\partial E^{\xi_-}_\mathrm{xc}[n_N]}{\partial \xi_-}\right|_{\xi_-=0} \right)\\
    &\equiv N \left.\dfrac{\partial E^{\xi}_\mathrm{xc}[n_N]}{\partial \xi}\right|_{\xi=0},
\end{split}
\end{equation}
which clearly shows that, as expected, $v^{\xi_+\rightarrow
0^+}_\mathrm{xc}(x)\neq v^{\xi_+=0}_\mathrm{xc}(x)$. A consequence of Eqs.~(\ref{eq:pot_infinity_constr}) and (\ref{eq:connection_weightderiv_DD}) is the constant shift $\Delta$ which manifests in the xc potential \textit{in the
region of the atom or molecule} as $\xi_+$ infinitesimally increases above zero, as shown in Fig.~\ref{Fig:EA_exact_atom} (Middle). The same occurs in the PPLB approach when the number of electrons within the system infinitesimally increases above an integer; see Fig.~\ref{Fig:PPLB}. Therefore, Eq.~(\ref{eq:connection_weightderiv_DD}) can be simplified as follows:
\begin{equation}\label{eq:shift_in_Nc-ens_xcpot_simp}
    \Delta  \int \mathrm{d}x\, n_N(x)=N \left.\dfrac{\partial E^{\xi}_\mathrm{xc}[n_N]}{\partial \xi}\right|_{\xi=0}, \nonumber
\end{equation}
thus leading to
\begin{equation}\label{eq:shift_in_Nc-ens_xcpot}
\begin{split}
    \Delta&=v^{\xi_+\rightarrow 0^+}_\mathrm{xc}(x)-v^{\xi_+=0}_\mathrm{xc}(x) \\ 
    &= \left.\dfrac{\delta E^{\xi_+}_\mathrm{xc}[n]}{\delta n(x)}\right|_{n=n_N,\xi_+ \rightarrow 0^+} - \left.\dfrac{\delta E^{\xi_+}_\mathrm{xc}[n]}{\delta n(x)}\right|_{n=n_N,\xi_+=0} \\
    &=\left.\dfrac{\partial E^{\xi}_\mathrm{xc}[n_N]}{\partial \xi}\right|_{\xi=0},
\end{split}
\end{equation}
which shows that, within the $N$-centered picture, the weight derivative $\Delta$ manifests in the exact xc potential as a derivative discontinuity, when switching from left to right ensembles. Equation~(\ref{eq:shift_in_Nc-ens_xcpot}), which is a key result, is the analogue for charged excitations of Levy's derivative discontinuity expression for neutral excitations (see Eq.~(29) in Ref.~\onlinecite{PhysRevA.52.R4313}; see also Ref.~\onlinecite{PhysRevA.90.042501}).

In summary, if we force the $N$-centered ensemble xc potential to asymptotically approach zero, we recover IP/EA theorems which are formally identical to those derived in PPLB. In this case, the exact KS HOMO and LUMO energies are weight-\textit{independent} and
\begin{align}
    I-A&=\varepsilon^{\xi_+\rightarrow 0^+}_{N+1}-\varepsilon^{\xi_-=0}_N \nonumber \\
    &=\varepsilon^{\xi_+\rightarrow 0^+}_{N+1}-\varepsilon^{\xi_+=0}_N \\
    &= \varepsilon^{\xi_+=0}_{N+1}+v^{\xi_+\rightarrow 0^+}_\mathrm{xc}(x)-v^{\xi_+=0}_\mathrm{xc}(x)-\varepsilon^{\xi_+=0}_N \nonumber \\
    &=\varepsilon_{N+1}-\varepsilon_{N}+\Delta. \nonumber
\end{align}
If we relax the constraint on the $N$-centered ensemble xc potential at infinite distance, we can still define a truly unique potential {\it via} the LZ shifting procedure $v^{\xi_+}_\mathrm{xc}(x)\rightarrow \overline{v}^{\xi_+}_\mathrm{xc}(x)= v^{\xi_+}_\mathrm{xc}(x)+\left . C^{\xi_+}[n] \right \vert_{n=n^{\xi^+}}$. In the latter case, the derivative discontinuity $\Delta$, which can be evaluated through an ensemble weight derivative, is moved away from the system. Indeed, according to Eqs.~(\ref{Eq:LZ}), (\ref{eq:connection_weightderiv_DD}), and (\ref{eq:shift_in_Nc-ens_xcpot_simp}), 
\begin{equation}
\begin{split}
    &\overline{v}^{\xi_+\rightarrow 0^+}_\mathrm{xc}(x)-\overline{v}^{\xi_+=0}_\mathrm{xc}(x) \\
    &=v^{\xi_+\rightarrow 0^+}_\mathrm{xc}(x)-v^{\xi_+=0}_\mathrm{xc}(x) \\
    &\quad+\dfrac{\int \mathrm{d}x \left[-v^{\xi_+\rightarrow 0^+}_\mathrm{xc}(x)+v^{\xi_+=0}_\mathrm{xc}(x)\right]n_{N}(x)}{\int \mathrm{d}x\, n_N(x)} \\
    &=v^{\xi_+\rightarrow 0^+}_\mathrm{xc}(x)-v^{\xi_+=0}_\mathrm{xc}(x)-\Delta,
    \end{split}
\end{equation}
which gives $\overline{v}^{\xi_+\rightarrow 0^+}_\mathrm{xc}(x)=\overline{v}^{\xi_+=0}_\mathrm{xc}(x)$ in the region of the atom or molecule and, according to Eq.~(\ref{eq:pot_infinity_constr}),
\begin{equation}
\begin{split}
    &\overline{v}^{\xi_+\rightarrow 0^+}_\mathrm{xc}(\left | x\right | \rightarrow \infty)
    -\overline{v}^{\xi_+=0}_\mathrm{xc}(\left | x\right | \rightarrow \infty)=-\Delta.
    \end{split}
\end{equation}

\section{Conclusion}\label{Sec:Conclusion}

The `$N$-centered' system corresponds to an ensemble Kohn-Sham (KS) system in which the electron number is fixed and integer. This constraint on the ensemble system permits the fundamental gap to be calculated from the KS gap, not through the usually unattainable `derivative discontinuity', $\Delta$, which manifests in the exact exchange-correlation (xc) potential as a discontinuous shift within the usual Perdew, Parr, Levy, and Balduz (PPLB) approach \cite{perdew1982density}, but instead via the ensemble-weight derivative of the $N$-centered xc energy. 

We modeled the $N$-centered system exactly in 1D in order to study the \textit{exact} xc energy and corresponding xc potential as a function of the ensemble weight ($\xi$). We found that the xc potential can always be (arbitrarily) forced to asymptotically approach zero infinitely far from the center of the system. We employed this constraint to prove analytically that $\Delta$ manifests within the exact xc potential of the $N$-centered ensemble as a derivative discontinuity, thus extending to charged excitations Levy's exact property of number-conserving excitations~\cite{PhysRevA.52.R4313}. We demonstrated numerically and analytically that the exact Levy-Zahariev (LZ) shift `relocates' this discontinuity to the periphery of the system -- a region with negligible importance to the system's observable properties.

We employed a standard 1D local density approximation (LDA) to the xc energy which completely neglects the ensemble weight dependence. We found, for our prototype system, that the LDA LZ-shifted KS energies were accurate for a significant range of ensemble weights including $\xi=0$ which corresponds to the regular $N$-electron KS system. 

Within the PPLB perspective the LDA is unreliable because it does not exhibit the required discontinuous shift in the xc potential and hence predicting the electron affinity (EA) results in a substantial error. Additionally, the LDA yields an inaccurate ionization potential (IP) owing to the usual self-interaction error. From an $N$-centered perspective, the LDA is much more reliable for the LZ-shifted highest occupied and lowest unoccupied KS energies; by design, the LDA's inability to yield the discontinuous shift in the xc potential is inconsequential. We find that for $\xi=0$ the LDA is remarkably accurate for the LZ-shifted KS energies; the IP and EA are inaccurate because the standard LDA is independent of the ensemble weight, $\xi$, as expected. Recent work combining finite (for introducing weight dependencies into the functional) and infinite (for reducing finite-size effects) uniform electron gas models have yielded substantial progress in the design of local ensemble functionals for neutral excitations~\cite{loos2020weight,marut2020weight}. This strategy can also be applied to $N$-centered ensembles to yield an approximation to the ensemble weight dependence of the xc energy. This offers the prospect of calculations which yield reliable fundamental gaps at a substantially lower cost than with popular hybrid functionals.

\acknowledgements{We thank Rex Godby at the University of York for computational resources and the European Theoretical Spectroscopy Facility's Correlation Team Meeting from which this collaboration was conceived.}

\end{document}